\documentstyle[12pt]{article}
\begin{document}
\baselineskip 3.9ex
\def\be{\begin{equation}}
\def\ee{\end{equation}}
\def\ba{\begin{array}{l}}
\def\ea{\end{array}}
\def\bea{\begin{eqnarray}}
\def\eea{\end{eqnarray}}
\def\no#1{{\tt   hep-th#1}}
\def\eq#1{(\ref{#1})}
\def\pgap{\vspace{1.5ex}}
\def\ggap{\vspace{10ex}}
\def\gap{\vspace{3ex}}
\def\del{\partial}
\def\o{{\cal O}}
\def\z{{\vec z}}
\def\re#1{{\bf #1}}
\def\av#1{{\langle  #1 \rangle}}
\def\S{{\cal S}}

\renewcommand\arraystretch{1.5}
\begin{flushright}
February 2000 \\
hep-th/0002202 \\
\end{flushright}
\begin{center}
\vspace{2 ex}
{\large{\bf Infra-red dynamics of D1-branes at the conifold}} \\
\vspace{3 ex}
Justin R. David \\
{\sl Department of Physics} \\
{\sl University of California, } \\
{\sl Santa Barbara, CA 93106, USA.} \\
\vspace{10ex}
\pretolerance=1000000
\bf ABSTRACT
\end{center}
\vspace{1ex}

We study the infra-red dynamics of D1-branes at the conifold. We show
using methods developed to study the infra-red dynamics of $(4,4)$
theories, the infra-red degrees of freedom of the 
$(2,2)$ theory of a single D1-brane at the conifold is that of a
linear dilaton with background charge of $\sqrt{2}$ and a compact
scalar. The gauge theory of $N$ D1-branes at the conifold is used to
formulate the matrix string in the conifold background.

\vfil
\hrule
\vspace{0.5 ex}
\leftline{justin@vulcan.physics.ucsb.edu}
\clearpage

\section{Introduction}

To explore the duality between large $N$ gauge theories 
and supergravity \cite{mal,witN,gpk} it is
important to study cases with less supersymmetry and theories which are
non-conformal \cite{imsy}.  
In this letter we study an example of such a theory.
This theory is obtained on D1-branes at a conifold singularity.
The conifold preserves $1/4$ of the supersymmetries of the full type
IIB string theory. The theory on the D1-brane is a supersymmetric
gauge theory in $1+1$ dimensions with $4$ supercharges.

We construct the supergravity solution of this
configuration. We investigate the decoupling limit and find the
domains of validity of supergravity description and the
super-Yang-Mills description. We see that the infra-red limit of
the super-Yang-Mills corresponds to matrix string theory in the 
background of the conifold. 

Thus the infra-red limit of the super-Yang-Mills on a 
single  D1-brane at
the conifold should correspond to world sheet of a  
fundamental string propagating in
the background of the conifold. 
We study the infra-red dynamics of the D1-brane gauge theory using the
methods developed for $(4,4)$ theories by \cite{ahaberk} following
\cite{wittconf} and\cite{verber}.
The theory on the D1-brane at the conifold has $(2,2)$ supersymmetry.
There is a 1-1 map from the moduli space of the Higgs branch of the
D1-brane gauge theory to the conifold.  The throat region of the Higgs
branch corresponds to the singularity at the origin of the conifold.
Though our theory has only $(2,2)$ supersymmetry most of the methods
developed by \cite{ahaberk} to study $(4,4)$ theories go through. 
Their method involves using the Coulomb variables to give an
effective description of the throat region of the Higgs branch.
In theories with $8$ supercharges the
Coulomb branch moduli space metric can receive correction only up to
1-loop. We do not have such facility for the case of $(2,2)$ theories.
Nevertheless scale invariance constraints the metric to a form
which enables us to extract the effective degrees of freedom.
The matching of the R-symmetries in the ultra-violet and the infra-red
works out just as in the case of $(4,4)$ theories.

Using these methods we are able to show that the throat region of the
Higgs branch in the infra-red is captured by a ${\cal N}=(2,2)$
superconformal field theory consisting of a linear dilaton with 
background charge $Q=\sqrt{2}$ and a compact scalar. This agrees with
the world sheet descriptions of strings at the conifold.

Then we consider $N$ D1-branes at the conifold. The infra-red limit of
this theory corresponds to matrix string theory in the background of
the conifold. 
The conifold provides a background in which the ultra-violet $U(1)_R$
symmetry is realized in the infra-red world sheet symmetry of the
matrix string theory. We construct the leading 
interaction in the form of the twist operator. We note that the
leading interaction is marginal.

The organization of this letter is as follows. In section 2 we study
the decoupling limit of the N D1-branes at the conifold and
investigate the domains of validity of the supergravity and the gauge
theory. Section 3 analyses the infra-red dynamics of the gauge theory
of a single D1-brane at the conifold. Section 4 formulates matrix
string theory in the background of the conifold. We conclude in
section 5 The appendix contains details of the
supergravity solution.

\section{Supergravity and the large $N$ limit of the D1-brane theory
at the conifold}

In this section we study the supergravity solution of $N$
D1-branes at a conifold singularity in the decoupling
limit. We consider the configuration in which the D1-branes are
aligned along the $x^1$ co-ordinate. The supergravity solution is
given by 
(The verification  of this solution is given in the Appendix.)
\bea
\label{sugrasoln}
ds^2 &=& f^{-1/2} (-dx_0^2 + dx_1^2) + f^{1/2} \left[ dr_1^2 +
r_1^2d\chi^2  \right. + dr_2^2\\ \nonumber
&+&  \frac{r_2^2 }{9}\left( d\psi +\cos\theta_1d\phi_1
+\cos\theta_2d\phi_2 \right)^2 +
\frac{r_2^2}{6} \left. \sum_{i=1}^{2} \left(
d\theta_i^2 + \sin^2\theta_id\phi_i^2 \right)  \right] \\
\nonumber
e^{(\Phi -\Phi_\infty)}& =& f^{1/2} \\ \nonumber
B_{01} &=& -\frac{1}{2} (f^{-1} -1) \\ \nonumber
f &=& 1+ \frac{N C g_s \alpha^{\prime 3}}{(r_1^2 + r_2^2)^3}
\eea
The co-ordinates transverse to the D1-brane are $r_1, \chi, r_2, \psi,
\theta_i, \phi_i$. $i$ runs from 1 to 2. 
$r_1$, $\chi$ are polar co-ordinates of  $R^2$. The remaining
coordinates parameterize the conifold. 
The angular part of the conifold is parametrized by $ \psi,
\theta_i, \phi_i$, the radial part is parametrized by $r_2$.
$(\theta_1,
\phi_1)$ and $(\theta_2, \phi_2)$ parameterizes $S^2\times S^2$ as polar
co-ordinates. $\psi\in [0, 4\pi]$ parameterizes the $U(1)$ fiber over
$S^2\times S^2$. $C$ is fixed by charge quantization. It is given by
$C= 864\pi^2/(16 + 15\pi)$. This configuration preserves $4$
supersymmetries out of the $32$ supersymmetries of Type IIB
supergravity.

We study the decoupling limit of this system as in \cite{imsy}. It
takes the form
\be
U_1=\frac{r_1}{\alpha^{\prime}}=\hbox{fixed}, \;\;
U_2=\frac{r_2}{\alpha^{\prime}}=\hbox{fixed}, \;\;
g^2_{YM} =\frac{1}{2\pi}\frac{g_s}{\alpha^{\prime}} =\hbox{fixed}, \;\;
\alpha^{\prime}\rightarrow 0
\ee
The metric and the dilaton of the supergravity solution in this limit
is given by
\bea
\label{nearsol}
\frac{ds^2}{\alpha^{\prime}} 
&=&  \frac{U^3}{g_{YM} \sqrt{2\pi NC} }
(-dx_0^2 + dx_1^2) + \frac{g_{YM}\sqrt{2\pi NC}}{U^3} \left[ dU_1^2 +
U_1^2d\chi^2 + dU_2^2 + \right.  \\ \nonumber
&+&  \frac{U_2^2 }{9}\left( d\psi +\cos\theta_1d\phi_1
+\cos\theta_2d\phi_2 \right)^2 +
\frac{U_2^2}{6} \left. \sum_{i=1}^{2} \left(
d\theta_i^2 + \sin^2\theta_id\phi_i^2 \right)  \right] \\ \nonumber
e^{\Phi} &=& \sqrt{\frac{8\pi^3NCg^6_{YM}}{U^6}}
\eea
where $U^2=U_1^2 + U_2^2$.

Let us now discuss the domains of validity of various descriptions of
this system. We will use the co-ordinate $U$ to set the 
energy scale at which we wish to look at the system. We will start
with the high energies. For $U\gg g_{YM}\sqrt{N}$ the super-Yang-Mills
perturbation theory can be trusted. 

To find out when the supergravity solution given in \eq{nearsol} 
can be trusted let us estimate the curvature of the solution. Using
the equation of motion an estimate of the curvature in string units is
given by
\bea
\alpha^{\prime}R&\sim& g^{U_1U_1}\partial_{U_1}\Phi\partial_{U_1}\Phi +
g^{U_2U_2}\partial_{U_2}\Phi\partial_{U_2}\Phi \\ \nonumber
&\sim& \frac{U}{g_{YM}\sqrt{N}}
\eea
To trust supergravity the curvature should be small. Furthermore we
need to ensure that the expansion in string coupling is valid. This is
requires $e^{\Phi}$ in \eq{nearsol} to be small. 
Thus the supergravity solution is valid for $g_{YM}N^{1/6}\ll U \ll
g_{YM}\sqrt{N}$. In addition to this condition, we must have
$U\ll U_2^{2/3}(Ng_{YM}^2)^{1/6}$. The latter condition arises from 
the fact that there is
a curvature singularity at $U_2=0$ \footnote{This point was raised by
G. Horowitz and N. Itzhaki}. 

In the region $U\ll g_{YM}N^{1/6}$ we can use S-duality to study the
solution. Performing S-duality on the near horizon solution in
\eq{nearsol} we obtain
\bea
\label{nearsdual}
\frac{ds^2}{\tilde{\alpha}^{\prime} } 
 &=&  \frac{U^6}{g_{YM}^4  4\pi ^2 NC  }
(-dx_0^2 + dx_1^2) + \frac{1}{2\pi g_{YM}^2} \left[ dU_1^2 +
U_1^2d\chi^2 + dU_2^2 +  \right. \\ \nonumber
&+&  \frac{U_2^2 }{9}\left( d\psi +\cos\theta_1d\phi_1
+\cos\theta_2d\phi_2 \right)^2 +
\frac{U_2^2}{6} \left. \sum_{i=1}^{2} \left(
d\theta_i^2 + \sin^2\theta_id\phi_i^2 \right)  \right] \\ \nonumber
e^{-\Phi} &=& \sqrt{\frac{8\pi^3NCg^6_{YM}}{U^6}}
\eea
where $\tilde{\alpha}^{\prime} = g_s\alpha^{\prime}$. The supergravity
solution in \eq{nearsdual} is the near horizon geometry of $N$
fundamental strings at the conifold. 
An estimate of the curvature in string units can be performed as
before, this gives the following
\be
\tilde{\alpha}^{\prime} R \sim \frac{g_{YM}^2}{U^2}
\ee
This shows that this S-dual supergravity description 
is valid for $U\gg g_{YM}$. As before we also have an additional
condition $\alpha^{\prime} U_2\gg 1$ so as to avoid the curvature
singularity at $U_2=0$.  For small $U$ we have no supergravity
description regardless of $N$. 

If one  is able to capture the infra-red behaviour of the
super-Yang-Mills on the D1-branes then it is clear that one obtains a
non-perturbative description of propagation of fundamental strings in
the background of the conifold. This is analogous to case of D1-branes
in flat space. The infra-red behaviour of super-Yang-Mills with gauge
group $U(N)$ and $16$ supercharges provides a nonperturbative
description of string theory \cite{motl,banks,dvv}.  
It is interesting to compare this also
with the  infra-red behaviour of
the D1/D5 system. There the infra-red dynamics of the D1-branes
captures the DLCQ of the little string theories in the Higgs branch.
The Coulomb branch conformal field theory gives a non-perturbative
description of srings in the background of Neveu-Schwarz 5-branes.

\section{Infra-red dynamics of a single D1-brane}

In this section we show that the
gauge theory of a single D1-brane at the 
conifold in the infra-red flows to
a superconformal field theory  
of a string in the background of the conifold.

\subsection{The gauge theory}

The gauge theory on the D1-brane at the conifold
consists of a $U(1)\times U(1)$ gauge
theory with $(2,2)$ supersymmetry in $1+1$ dimensions \cite{klebwit}. 
The matter content of this theory consists
of two sets of chiral multiplet, 
$A_i$,  and $B_i$ with $i=1,2$. The $A$'s and $B$'s are charged as
$(1,-1)$ and $(-1,1)$ respectively. The diagonal $U(1)$ decouples. It
corresponds to the free $U(1)$ gauge multiplet on the single D1-brane.
The $2$ scalars of this gauge multiplet represent motion of the
D1-brane along $r_1$ and $\chi$. 
Under the relative $U(1)$ the $A$'s 
 have charge $+1$ and the $B$'s have charge $-1$. 
 The D-term of the $U(1)$ vector multiplet is given by
 \be
 D= |A_1|^2 + |A_2|^2 - |B_1|^2 - |B_2|^2
 \ee
 We consider the case in which both the Fayet-Iliopoulos term and the
 theta term in the Lagrangian are set to zero. The conifold is
 realized as the moduli space of vacuum of the Higgs branch of this
 theory. Setting the D-term to zero and dividing by the gauge group
 $U(1)$ realizes the conifold. The complex coordinates of the conifold
 are given by
 \be
z_1 =A_1B_1, \;\;\; z_2=A_2B_2, \;\;\; z_3=A_1B_2, \;\;\; z_4= A_2B_1
\ee
with
\be
\label{conifold}
z_1z_2-z_3z_4=0
\ee
Therefore the infrared theory is a  superconformal field theory with
the conifold as its target space. The central charge in the Higgs
branch is given by counting the gauge invariant degrees of freedom.
This is seen to be $9$.
To isolate the description of the conifold at the
singularity we describe the conifold as follows. 

If point $(a_1, a_2, a_3, a_4)$ satisfies \eq{conifold}, and $a_i\neq 0$
then one can obtain another solution which is given by
$\sigma^{1/2} (a_1, a_2, a_3, a_4)$. Here $\sigma$ is a complex
number. This particular scaling is chosen so that the complex polynomial
describing the conifold in \eq{conifold} is homogeneous of degree
$1$. Therefore the conifold can be described by the space
\be
\sigma\times ( z_1z_2-z_3z_4=0)/\sigma
\ee
where the space $(z_1z_2-z_3z_4)/\sigma$ is $2$ complex dimensional
hypersurface $z_1z_2-z_3z_4=0$ in the $3$ complex dimensional weighted
projective space
$WCP^3_{\frac{1}{2},\frac{1}{2},\frac{1}{2},\frac{1}{2}}$.

The central charge of the 
${\cal N} =(2,2)$ supeconformal field theory on the hypersurface in
the weighted projective space is zero. 
Thus it does not contain any degrees of
freedom. The entire central charge $9$ of the superconformal field
theory of the conifold thus resides on the superconformal field 
theory on the one dimensional complex space parametrized by $\sigma$. 
It is clear this space is endowed with a nontrivial metric. 
To obtain this metric we examine the theory at the origin of the Higgs
branch. At the origin of the Higgs branch there is a `singularity' as
the Fayet-Iliopoulos term and the theta term are set to zero. This
corresponds to the  $z_i=0$ point on the conifold. In the infra-red we
can describe the throat region of the Higgs branch using the Coulomb
variables. This is because in the infra-red, the vector multiplet is
an auxiliary field. The kinetic terms of the vector multiplet
decouples as they are irrelevant. Then the Higgs fields can be written
in terms of the vector multiplet using equations of motion.
This method of describing the Higgs branch using the Coulomb variables
was done in \cite{ahaberk} of $N=(4,4)$ gauge theories. We obtain the
metric on the complex line parametrized by $\sigma$ 
by appealing to the description of the Higgs branch in
terms of the Coulomb branch.

To proceed with the analysis we write down the Lagrangian for the
relative $U(1)$. We follow the convention of \cite{witphase} but work
with Euclidean world sheet metric. We set $y^0=-iy^2$ in the formulae
of \cite{witphase}.
\bea
L&=& L_{\rm matter} + L_{\rm gauge} \\ \nonumber
L_{\rm matter} &=& \int d^2y D_\mu \bar{A}_i D^\mu A_i  
+D_\mu \bar{B}_i D^\mu B_i + 2\sigma\bar{\sigma}\bar{A}_iA_i 
+ 2\sigma\bar{\sigma}\bar{B}_iB_i  \\ \nonumber
&-& i\bar{\psi}^A_{-i} ( D_1 + iD_2)\psi_{-i}^A
-i\bar{\psi}^B_{-i} ( D_1 + iD_2)\psi_{-i}^B \\ \nonumber
&-&i\bar{\psi}^A_{+i} ( D_1 -i D_2)\psi_{+i}^A
-i\bar{\psi}^B_{+i} ( D_1 - iD_2)\psi_{+i}^B \\ \nonumber
&+& \sqrt{2} (\bar{\sigma} \bar{\psi}_{+i}^A\psi_{-i}^A + \sigma
\bar{\psi}^A_{-i} \psi^A_{+i} )
- \sqrt{2} (\bar{\sigma} \bar{\psi}_{+i}^B\psi_{-i}^B + \sigma
\bar{\psi}^B_{-i} \psi^B_{+i} ) \\ \nonumber
&+&i\sqrt{2} \bar{A}_i (\psi_{-i}^A\lambda_{+} - \psi_{+i}^A\lambda_{-}
)
+i\sqrt{2} A_i (\bar{\lambda}_{-} \bar{\psi}^A_{+i} -\bar{\lambda}_{+}
\bar{\psi}^{A}_{-i}) \\ \nonumber
&+&i\sqrt{2} \bar{B}_i (\psi_{-i}^B\lambda_{+} - \psi_{+i}^B\lambda_{-}
)
+i\sqrt{2} B_i (\bar{\lambda}_{-} \bar{\psi}^B_{+i} -\bar{\lambda}_{+}
\bar{\psi}^{B}_{-i}) \\ \nonumber
&-& D\bar{A}_iA_i + D \bar{B}_iB_i \\ \nonumber
L_{\rm gauge} &=& \frac{1}{g_{YM}^2} \int d^2y \left( \frac{1}{2}
F_{01}^2 -\frac{1}{2} D^2 \right. \\ \nonumber
&-& \left. i\bar{\lambda}_{+}(\partial_1-i\partial_2)\lambda_+
-i\bar{\lambda}_{-}(\partial_1+i\partial_2)\lambda_  -
+\partial_\mu\bar{\sigma}\partial^\mu\sigma \right)
\eea
The superpotential on a single D1-brane is zero, therefore we have set
the $F$-terms to zero. The vector multiplet consists of fields
$F_{01}, \sigma, \lambda_{+},\lambda_{-}$. $\sigma$ is a complex
scalar corresponding to the components of a $4$ dimensional gauge
field  along  $2$ compact directions. The gauginos $\lambda_{+},
\lambda_{-}$ are complex Weyl fermions.
The super partners of the chiral multiplet $A_i, B_i$ are 
$\psi^{A}_{i+,-},\psi^{A}_{i+,-}$ respectively. They are also complex
Weyl fermions in $2$ dimensions.

The scaling dimension of the Yang-Mills coupling is $1$. Therefore in
 the infra-red the coupling tends to $\infty$. 
 This ensures that  the kinetic term for the vector multiplet
 $L_{\rm gauge}$ decouples in the infra-red. In the Higgs branch
 the scaling dimension of the scalars $A_i, B_i$ is zero, 
 and its superpartners have dimension $1/2$.
 The scalars $\sigma$ and the gauge boson have scaling 
 dimension $1$. Its superpartners have scaling dimension $3/2$. This
 is more evidence that the operators in $L_{\rm gauge}$ are
 irrelevant. Therefore in the infra-red, the Lagrangian is restricted
 to only $L_{\rm matter}$. 

 We now can integrate over the auxiliary vector multiplet in
 $L_{\rm matter}$. This forces the D-term to be set to zero and one
 obtains the Higgs branch as a 
 ${\cal N}=(2,2)$ superconformal field theory over the conifold. 
 However in order to  describe the theory near the singularity
 we will follow the method of \cite{ahaberk}. Here the vector
 multiplets are regarded as composite operators on the Higgs branch.  
They are roughly given by
\be
\label{vector}
\sigma = \frac{1}{\sqrt{2}} \frac{ \bar{\psi}_{-i}^B\psi_{+i}^B -
\bar{\psi}_{-i}^A\psi_{+i}^A }{\bar{A}_iA_i + \bar{B}_iB }
\ee
This amounts to integrating out the chiral multiplets in 
$L_{\rm matter}$. This was argued in \cite{ahaberk} to be valid at
large values of $\sigma$. From \eq{vector} we see that is valid
roughly for small values of the chiral multiplets $A_i, B_i$. Thus by
large values of $\sigma$ we are probing the singularities of the Higgs
branch. We will discuss the systematics of the expansion as we perform
the 1-loop computation.

\subsection{The one-loop calculation}

The terms in the Lagrangian which are relevant for the 1-loop
calculation are
\bea
L &=& \int d^2y \partial_\mu \bar{A}_i \partial^\mu A_i  
+\partial_\mu \bar{B}_i \partial^\mu B_i + 
2\sigma\bar{\sigma}\bar{A}_iA_i 
+ 2\sigma\bar{\sigma}\bar{B}_iB_i  \\ \nonumber
&-& i\bar{\psi}^A_{-i} ( \partial_1 + i\partial_2)\psi_{-i}^A
-i\bar{\psi}^B_{-i} ( \partial_1 + i\partial_2)\psi_{-i}^B \\
\nonumber
&-&i\bar{\psi}^A_{+i} ( \partial_1 - i\partial_2)\psi_{+i}^A
-i\bar{\psi}^B_{+i} ( \partial_1 - i\partial_2)\psi_{+i}^B \\ \nonumber
&+& \sqrt{2} (\bar{\sigma} \bar{\psi}_{+i}^A\psi_{-i}^A + \sigma
\bar{\psi}^A_{-i} \psi^A_{+i} )
- \sqrt{2} (\bar{\sigma} \bar{\psi}_{+i}^B\psi_{-i}^B + \sigma
\bar{\psi}^B_{-i} \psi^B_{+i} ) 
\eea

Integrating out the chiral multiplets to 1-loop 
gives the following terms in the action
\bea
S_{{\rm 1-loop}} &=&
-4\ln \left[ \hbox{det}(-\Box + 2\bar{\sigma}\sigma) \right]
+ 2\ln \left[ \hbox{det}
\left(
\begin{array}{cc}
-i(\partial_1 + i\partial_2) & +\sqrt{2}\sigma  \\
+\sqrt{2} \bar{\sigma} &i(\partial_1-i\partial_2)
\end{array} \right)
\right] \\ \nonumber
&+& 2 \ln \left[ \hbox{det}
\left(
\begin{array}{cc}
-i(\partial_1 + i\partial_2) & -\sqrt{2}\sigma  \\
-\sqrt{2} \bar{\sigma} &i(\partial_1-i\partial_2)
\end{array} \right)
\right] \\ \nonumber
\eea
On simplification one gets
\bea
\label{determinant}
S_{\rm 1-loop} &=&
\hbox{Tr}\ln \left[ 1 +
\left(
\begin{array}{cc}
0 & 
\frac{1}{-\Box + 2\sigma\bar{\sigma}}
[-\sqrt{2}(\partial_1+i\partial_2)]\sigma \\
\frac{1}{-\Box + 2\sigma\bar{\sigma}}
[\sqrt{2}(\partial_1-i\partial_2)]\bar{\sigma} & 0
\end{array}
\right) \right] \\ \nonumber
&+& \hbox{Tr}\ln \left[ 1 +
\left(
\begin{array}{cc}
0 & \frac{1}{-\Box + 2\sigma\bar{\sigma}}
[\sqrt{2}(\partial_1+i\partial_2)]\sigma \\
-\frac{1}{-\Box + 2\sigma\bar{\sigma}}
[i\sqrt{2}(\partial_1-i\partial_2)]\bar{\sigma} & 0
\end{array}
\right) \right]
\eea
From \eq{determinant} it is clear that the expansion parameter is
$\frac{d\sigma}{\sigma^2}$. We expect this expansion to be valid
for $d\sigma \ll \sigma^2$. Thus, we obtain a good description of the
Higgs branch in terms of the Coulomb variables for large $\sigma$,
which according to \eq{vector} 
corresponds to regions near the singularity.
After further simplification the 
leading order in the velocity expansion is given by
\bea
S_{\rm 1-loop} &=&
4\int d^2 y 
(\partial_1+i\partial_2)\sigma(\partial_1-i\partial_2)\bar{\sigma}
\hbox{Tr} \left[ \frac{1}{(-\Box + 2\sigma\bar{\sigma})^2}\right] \\
\nonumber
&=& 4\int d^2 y 
(\partial_1+i\partial_2)\sigma(\partial_1-i\partial_2)\bar{\sigma}
\int\frac{d^2k}{4\pi^2} \frac{1}{(k^2 + 2 \sigma\bar{\sigma})^2}
\eea
$\sigma$ is a complex scalar which represents the 2 coordinates of the
coulomb branch. Writing these in polar coordinates 
and performing the integration we obtain the
following metric on the moduli space.
\be
\label{modulimetric}
ds^2 = \frac{dr^2}{2\pi r^2} + \frac{d\theta^2}{2\pi}
\ee
There is also a torsion given by
\be
B_{r\theta} = \frac{1}{2\pi r}
\ee
The torsion is a pure gauge term. 
The space is topologically $R\times S^1$. This does not
have any nontrivial closed 2-cycles. Thus there is no obstacle for
gauging away the torsion. The gauge transformation is given by 
\be
B_{r\theta} = \partial_r\Lambda_{\theta} -\partial_{\theta} \Lambda_r
\ee
Setting $\Lambda_r=0$ gives $\Lambda_{\theta} = \ln r/2\pi$.
We have performed only a 1-loop calculation for the moduli
space metric. 
The moduli space metric in \eq{modulimetric} 
can be argued to be of the form given by the 1-loop result by
scale invariance.  
$\sigma$ is a scalar with dimension $1$ in the ultra-violet
gauge theory. The only scale invariant term one can write for the
moduli space metric is $d\sigma d\bar{\sigma}/\sigma\bar{\sigma}$.
We have determined the coefficient by the 1-loop calculation. 
A similar 1-loop calculation was done for the $(4,4)$ relevant for
the D1/D5 system in \cite{dps}. In
$(4,4)$ gauge theories a non-renormalization theorem determines the
moduli space metric by the 1-loop result \cite{gates,dps,seidia}. 
For the $(2,2)$ case there
is no such theorem, but conformal invariance constraints the metric
up to a numerical coefficient. To determine the infra-red degrees of
freedom, we do not need this coefficient.

\subsection{The infra-red degrees of freedom}

It is clear from the moduli space metric in \eq{modulimetric} that in
the infra-red the dimension of the field $r$ is not determined. 
We define a scalar field $\phi$ as $r= e^{-\phi/2}$. Then the kinetic
term for $\phi$ is just that of a free field. 
The field $\phi$ can behave like a linear dilaton.
The dimension of $r$ is
specified only if one knows the background charge of the linear
dilaton $\phi$. We determine the background charge by requiring that
the central charge of the infra-red ${\cal N}=(2,2)$ superconformal
field theory to be $9$
which as we saw before was the central charge of the Higgs branch.
We will justify this using R-symmetries in the section 3.4.

The bosonic fields 
capturing the infra-red dynamics are the linear dilaton
$\phi$, the compact scalar $\theta$. The radius of the compact scalar
is 2. This is the value determined by the 1-loop calculation.
The bosonic part of the action is given by
\be
L = \frac{1}{8\pi} \int d^2y \sqrt{g} ( g^{\mu\nu}\partial_\mu \phi
\partial_\nu \phi - QR\phi + g^{\mu\nu}\partial_\mu\theta
\partial_\nu\theta)
\ee 
where $R$ is the world sheet curvature and $Q$ is the background
charge of the linear dilaton. We have redefined $\theta$ so that we
can use the $\alpha' =2$ convention.
The superpartners of $\phi$ and $\theta$
are free fermions as the curvature of the moduli space $R_{r\theta r
\theta}$ is zero. 

Now we can evaluate the central charge. The total central charge of
the infra-red superconformal field theory is
\be
c= 3/2 + 3Q^2 + 3/2
\ee
Demanding that the central charge be $9$ gives $Q^2=2$. The sign of 
the background charge is determined by requiring that the singularity
is at the strong coupling region.This fixes the background charge to
be $Q=\sqrt{2}$. At this point we mention that in the $(4,4)$ case the
sign was determined by the fact that the conformal field theory of the
Higgs branch and the conformal field theory of the Coulomb branch
could be considered as two different subalgebras of the large ${\cal
N}=(4,4)$ superconformal algebra.

We note that demanding that the central charge be $9$ does not
determine the infra-red conformal field theory completely. The central
charge of a conformal field theory is unchanged if it is deformed by
a marginal operator. One obvious marginal deformation is the radius
of the compact scalar. Thus we cannot determine the radius of the
compact scalar in the infra-red limit. We determined the radius only
using a 1-loop calculation. This certainly can change in the infra-red
limit.

To be explicit we write down the holomorphic generators of the 
infra-red ${\cal N} =(2,2)$ superconformal field theory. 
\bea
\bar{G} &=& \psi\partial_z \bar{X} + \partial_z\psi \\ \nonumber
G &=& \bar{\psi}\partial_z X + \partial_z\bar{\psi} \\ \nonumber
J_R &=& -\psi\bar{\psi} + i \sqrt{2}\partial_z X^2 \\ \nonumber
T &=& -\partial_z\bar{X}\partial_z X  
- \frac{1}{\sqrt{2}}\partial_z^2 X^1
 + \frac{1}{2} (-\partial_z\psi \bar{\psi}
+ \psi\partial_z\bar{\psi})
\eea
where 
\be
X= \frac{ X^1 + iX^2}{\sqrt{2}}, \;\;\; 
\bar{X} =\frac{X^1-iX^2}{\sqrt{2}}, \;\;\; 
\psi=\frac{\psi^1 + i\psi^2}{\sqrt{2}} \;\;\;
\bar{\psi} = \frac{\psi^1-i\psi^2}{\sqrt{2}}. 
\ee
The field $X^1$ corresponds to the linear dilaton $\phi$ and $X^2$
corresponds to the compact scalar $\theta$. The fermions $\psi^1,
\psi^2$ are the superpartners of $X^1$ and $X^2$ respectively.
There is a similar set of anti-holomorphic generators.

\subsection{Comparison of the R-symmetries}

We now compare the R-symmetries of the ultra-violet and the infra-red
theory and show that our identification of the infra-red degrees of
freedom is justified. 

In the D1-brane gauge theory there is a $U(1)_L\times U(1)_R$
R-symmetry. The fields and their charges under $U(1)_L$ are 
as follows \cite{witphase}, 
$(\psi_{+i}^A,\psi_{+i}^B, e^{i\theta}, \lambda_{-})$ have
charges $(-1,-1, 1, 1)$. The rest of the fields are uncharged under
$U(1)_L$. The fields 
$(\psi_{-i}^A,\psi_{-i}^B, e^{i\theta}, \lambda_{+})$ have charges
$(-1,-1, -1, 1)$ under $U(1)_R$. 
The rest of the fields are uncharged under $U(1)_R$. 
The infra-red behaviour of state localized far along the Higgs branch
is approximately free.
In a ${\cal N} =(2,2)$  
free super conformal field theory with a flat metric 
the R-symmetry does not act on the bosons. The $U(1)_L$ and $U(1)_R$ 
does not act on the bosons of the chiral multiplet $A_i$  and $B_i$.
Thus it is natural to identify the R-symmetry of the D1-brane gauge
theory with that of the ${\cal N} =(2,2)$ superconformal field
theory of the Higgs branch. As the ${\cal N} =(2,2)$ superconformal
algebra relates the R-symmetry to the central charge we are justified
in requiring that the central charge of the infra-red superconformal
field theory of the Higgs branch be $9$.

The Coulomb branch is parametrized by the bosons $|\sigma|,
e^{i\theta}$ As $e^{i\theta}$ is charged under $U(1)_L\times U(1)_R$
this cannot be the R-symmetry of the conformal field theory of the
Coulomb branch. It must be as the theory flows to the infra-red on the
Coulomb branch an R-symmetry is developed. This is similar to the
case of $(4,4)$ gauge theories \cite{wittconf}. In these theories
the $SU(2)_R$ symmetry of the gauge theory in the ultra-violet is a
candidate for the R-symmetry of the conformal field theory of the
Coulomb branch. This symmetry is enhanced to $SU(2)\times SU(2)$ as
the theory flows to the infra-red in the Coulomb branch. 

Let us now examine from the infra-red degrees of freedom of the Higgs
branch in the throat region whether the R-symmetry acts similar to 
$U(1)_L\times U(1)_R$. Let us focus on the holomorphic part, the
anti-holomorphic part follows similarly. The field 
$(e^{iX^2/\sqrt{2}}, \psi)$ are charged as $(1,1)$. This is what is
expected under the identification of $U(1)_L\times U(1)_R$ 
as the R-symmetry of the conformal field theory of the Higgs branch. 

\subsection{Fundamental strings at the conifold}

We have seen in section 2 that using arguments of \cite{imsy} that the
infra-red theory of the D1-brane at the conifold should correspond to
that of the world sheet of fundamental strings in the background of
the conifold. We wish to compare the infra-red theory obtained with
what is known about string propagation at the conifold.

String propagation at singularities have been studied recently in a
series of works \cite{givkutpel,givkut1,givkut2}. 
It is seen from these works that string
propagation at the conifold is described by a linear dilaton theory
with back ground charge $Q=\sqrt{2}$ and a compact scalar. The
effective degrees of freedom of the infra-red D1-brane gauge theory
precisely matches with this.

String propagation at the resolved conifold 
$z_1^2 + z_2^2 + z_3^2 + z_4^2 = \mu$, has been discussed in
\cite{mukvafa,goshvafa,oogvafa}. 
In these series of works it was argued that the world sheet
theory was  given by a ${\cal N}=(2,2)$ $SL(2,R)/U(1)$ Kazama-Suzuki
model at level 3. The $SL(2,R)/U(1)$ model away from the origin
consists of a linear dilaton with $Q=\sqrt{2}$ and a compact scalar
at the self dual radius \cite{gsw,witblack}. This also agrees with the
infra-red degrees of freedom of the D1-brane gauge theory.

\section{Matrix String theory in a Conifold background}

In this section we use the Lagrangian of $N$ D1-branes in a conifold
background to formulate matrix string theory in this background. 
To formulate matrix string theory we need that the spatial coordinate
of the two dimensional Yang-Mills to be compact \cite{motl,banks,dvv}.
To the knowledge of the author, matrix string theory has not been
formulated in a background with 8 supersymmetries. 

The Lagrangian of $N$ D1-branes at a conifold is that constructed in
\cite{klebwit} dimensionally reduced to two dimensions.
It consists of a $U(N)\times U(N)^{\prime}$ gauge theory with $(2,2)$
supersymmetry. We will use $d=4$ $N=1$ supersymmetry nomenclature to
classify our fields. 
There are two gauge multiplets corresponding to the two gauge groups.
The bosonic fields of the gauge multiplet consist of two bosons
transforming in the adjoint of the corresponding gauge group. The
fermions of the gauge multiplet are complex Weyl fermions in two
dimensions. They are spinors of $SO(2)$ the symmetry in the transverse
directions parametrized by $r_1$ and $\chi$. This can be seen from the
fact that they arise from dimensional reduction of 
complex Weyl fermions of $4$ dimension. They transform in the adjoint
representation of the gauge group.
We list the fields of the gauge multiplets below.
\bea
\hbox{Bosons} \;\;\;
A_\mu ,A^{\prime}_\mu , X_a, X^{\prime}_a \\ \nonumber
\hbox{Fermions}\;\;\; \lambda_{+}, \lambda_{-}, \lambda^{\prime}_{+}
\lambda^{\prime}_{-}
\eea
where $a=1,2$. The primes  over the field variables 
indicate that they transform under the gauge group $U(N)^{\prime}$.
There are 4 chiral multiplets arranged in two sets $A_i$ and $B_i$,
$i=1,2$. The $A_i$ transform as 
$U(N)\times \overline{U(N)^{\prime}}$, while the
$B_i$ transform as $\overline{U(N)}\times U(N)^{\prime}$. 
We use the the capital
$A$'s and $B$'s to indicate the superfields as well as the bosonic
component. The fields of the chiral multiplets are
\bea
\hbox{Superfield}\;\; A_i \;\;\;\; A_i, \psi_{-i}^A \psi_{+i}^A \\
\nonumber
\hbox{Superfield}\;\;  B_i \;\;\;\; B_i, \psi_{-i}^B \psi_{+i}^B 
\eea
The Lagrangian has a superpotential given by
\be
W= \frac{1}{2} \epsilon^{ij}\epsilon^{kl}\hbox{Tr} A_iB_kA_jB_l
\ee
We choose the gauge coupling of the two gauge groups to be identical.
The bosonic potential is given by
\bea
U&=& g_{YM}^2\sum_i\left|\frac{\partial W}{\partial A_i} \right|^2
+g_{YM}^2\sum_i\left|\frac{\partial W}{\partial B_i} \right|^2 +
\frac{1}{2g_{YM}^2} \hbox{Tr} D^2 +
\frac{1}{2g_{YM}^2} \hbox{Tr} D^{\prime 2} \\ \nonumber
&+& A_i^* (X_1^2 + X_2^2) A_i +  
+ A_i^* (X_1^{\prime 2} + X_2^{\prime 2}) A_i   
+ B_i^* (X_1^2 + X_2^2) B_i   
+ B_i^* (X_1^{\prime 2} + X_2^{\prime 2}) B_i  \\ \nonumber
&+&\frac{1}{2\tilde{g}_{YM}^2}[X_1, X_2]^2 + 
\frac{1}{2g_{YM}^2}[X^{\prime}_1, X^{\prime}_2]^2
\eea

Perturbative Type IIA string theory is realized out of the matrix
description in the $g_{YM}\rightarrow\infty$. 
From the superpotential we see that this limit selects out a vacuum. 
One such vacuum is in which all the $A$'s and $B$'s are diagonal. 
We analyze the theory around this vacuum.
The gauge group is broken
down to $U(1)^N$ in this vacuum. Each of the  $U(1)$ corresponds to
the center of mass $U(1)$ for the single D1-brane considered in
section 3 The Weyl group of $U(N)\times U(N)^{\prime}$
acts on the vacuum as
\bea
A\rightarrow SAS^{\prime\dagger} \\ \nonumber
B\rightarrow S^{\prime}B S^{\dagger}
\eea
We see that the Weyl group of $U(N)\times U(N)^{\prime}$ 
transform $A$ and $B$
to values which are diagonal with the entries permuted only if $S$ and
$S^{\prime}$ 
are the same element of the Weyl group. Such transformations
takes one vacuum to another. The gauge invariant vacuum is
given by identifying these. The $D$ terms  for the relative $U(1)$ for
each of the D1-brane reduce to
\bea
|A_1^m|^2 + |A_2^m|^2 - |B_1^m|^2 -|B_2^m|^2 = 0
\eea
where $m =1,\ldots N$. $N$ copies of the conifold is realized as the
moduli space of vacuum. The complex coordinates of the $N$ copies of
the conifold  are  given by
\be
\label{fold}
z_1^m = A_1^mB_1^m, \;\;\; z_2^m = A_2^mB_2^m, \;\;\; z_3^m =
A_1^mB_2^m, \;\;\; z_4=A_2^mB_1^m
\ee

Thus the conformal field theory that describes the infra-red limit of
this gauge theory is a sigma model on the orbifold target space
\be
\frac{(R^2\times C)^N}{S(N)}
\ee
where $S(N)$ is the symmetric group and $C$ stands for the conifold.
The $R^2$ refers to the transverse directions parametrized by $r_1$
and $\chi$. These arise from the values of scalars $X_a^m$ 
corresponding to the diagonal $U(1)$'s.
The question of finding backgrounds for matrix
string theory where an $U(1)_R$ is present in the ultraviolet gauge
theory which appears as the world sheet $U(1)_R$ was raised recently 
in \cite{eva}. 
In the ultraviolet of this matrix theory, there is a superpotential.
Therefore the chiral multiplets are charged under the $U(1)_R$ in the
ultrviolet. Thus this R-symmetry cannot be the $U(1)_R$ of the world
sheet theory in the infrared {\footnote{The author thanks E.
Silverstein and Y. S. Song 
for pointing out that for this matrix theory too, the
question in \cite{eva} is unresolved, correcting the erroneous
conclusion in the earlier draft.}}.

Using the results of section 3, the conformal field theory on the
conifold  near the singularity  that captures the infra-red limit of
the gauge theory is given by the orbifold 
\be
\label{ocft}
\frac{(R^2\times R_\phi\times S^1)^N}{S(N)}
\ee
where $R_\phi$ stands for the linear dilaton with background charge
$Q=\sqrt{2}$ and $S^1$ refers to the compact scalar. At this point let
us examine the domain of validity of the super conformal field theory
on the orbifold \eq{ocft}. The super conformal field theory on the
orbifold \eq{ocft} is valid when $g_{YM}\rightarrow\infty$ and
$z\rightarrow 0$. Here $z$ stands for all the co-ordinates in
\eq{fold}. Now, it is important that there exists a domain in these
limits that
the mass of the off-diagonal
chiral multiplets  can be neglected. The mass of the
off-diagonal chiral multiplets roughly goes as $g_{YM}^2 z^2$. Thus,
the super conformal field theory on the orbifold is valid in the
limits $g_{YM}\rightarrow\infty$, $z \rightarrow 0$ and $g_{YM}^2
z^2 \rightarrow\infty$.

We would like to construct the leading interaction vertex represented
by the twist operator corresponding to the $Z_2$ conjugacy class of
the permutation group. For this we focus on the orbifold
\be
\frac{(R^2\times R_\phi\times S^1)^2}{Z_2}
\ee
Going over to center of mass and relative coordinate, conformal field
theory on the following target space is realized
\be
(R^2\times R_\phi\times S^1)\times \frac{(R^3\times S^1)}{Z_2}
\ee
Here the linear dilaton $R_\phi$ has background charge
$Q=\sqrt{2}\times\sqrt{2}$. From the fact that the orbifold is a
$(R^3\times S^1)/Z_2$, the interaction vertex is represented by the
twist operator that is marginal. 
This is unlike the case of D1-branes in flat space where the leading
interaction was irrelevant \cite{dvv}. 
The fact that there is a marginal operator in the infrared in this
case is puzzling. This
results perhaps from the fact that the theory is strongly coupled
at the singularity.
The issue of whether this
marginal operator is turned on or not in the infrared theory is
important. If the operator is turned on there is no weak coupling
limit and persumably the
infrared behaviour does not look like a perturbative
matrix string theory. It is important to resolve this issue further.

\section{Conclusions}

We have used methods developed for the analysis of 
infra-red dynamics of $(4,4)$ gauge theories to study the infra-red
dynamics of the $(2,2)$ gauge theory on a D1-brane at the conifold.
We showed that the infra-red dynamics is captured by a ${\cal N}
=(2,2)$ superconformal field theory consisting of a linear dilaton
with background charge $Q=\sqrt{2}$ and a compact scalar. This agreed
with the expectation that the infra-red theory should correspond to
that of a fundamental string at the conifold.
We mention that these methods can be used to analyze infra-red
dynamics of $(2,2)$ theories with one dimensional Coulomb branch.

The Lagrangian of $N$ D1-branes at the conifold was used to formulate
matrix string theory on this background. We note that the leading
interaction represented by the twist operator in this case is marginal
unlike the case of D1-branes in flat space.

\vspace{1cm}

\noindent
{\bf Acknowledgements}

The author acknowledges useful discussions with 
Nissan Itzhaki, Kiril Krasnov, Juan Maldacena and Ashoke Sen. 
He thanks Rajesh Gopakumar, 
Gary Horowitz and Joe Polchinksi for
a careful reading of the manuscript, useful comments and
discussions. He also
acknowledges Avinash Dhar, Gautam Mandal and Spenta Wadia for a
careful reading of the manuscript, comments and advice.
The work of the author is supported by NSF grant PHY97-22022.

\appendix
\section{The supergravity solution}

We follow the following 
strategy to verify the supergravity solution in \eq{sugrasoln}.
We convert the solution given in \eq{sugrasoln} into the Einstein
metric and into the conventions of \cite{duff}. In these conventions
we make the following ansatz for the supergravity solution.
\bea
\label{sugraein}
ds^2 &=& f^{-3/4} (-dx_0^2 + dx_1^2) + f^{1/4}  \left[ dr_1^2 +
r_1^2d\chi^2  \right. + dr_2^2\\ \nonumber
&+&  \frac{r_2^2 }{9}\left( d\psi +\cos\theta_1d\phi_1
+\cos\theta_2d\phi_2 \right)^2 +
\frac{r_2^2}{6} \left. \sum_{i=1}^{2} \left(
d\theta_i^2 + \sin^2\theta_id\phi_i^2 \right)  \right] \\
\nonumber
e^{\Phi }& =& f^{1/2} \\ \nonumber
B_{01} &=& - f^{-1}  \\ \nonumber
\eea
where $f$ is an unknown function. We then 
show that the equations of motion
of the various field just reduce to the Laplacian for the function $f$
in the coordinates transverse to the D1-brane.
 
We first substitute this ansatz in the dilaton equation.
The dilaton eqation in the convention of \cite{duff} is given by 
\bea
\label{dilaton}
\partial_{MN}(\sqrt{-g}g^{MN}\partial_N \Phi) -\frac{1}{24}
\sqrt{-g}e^{\Phi}H^2 \\ \nonumber
= -\frac{\kappa^2T_2}{2}\int d^2\xi\sqrt{-\gamma} \gamma^{ij}
\partial_i X^M \partial_j X^N g_{MN} e^{-\Phi/2}\delta^{10}(x-X)
\eea
Substituting the values of the field given in \eq{sugraein} in the
static gauge we find that the dilaton equation \eq{dilaton} reduces to
\be
\label{laplac}
\frac{1}{2f} \partial_{r_1}(K\partial_{r_1}f) + \frac{1}{2f}
\partial_{r_2} (K\partial_{r_2} f) = 
-T_2\kappa^2 K \frac{ 108 }{(4\pi)^3 f}
\frac{\delta(r_1)}{r_1}\frac{\delta(r_2)}{r_2^5}
\ee
where 
\be
K= \frac{r_1r_2^5}{108} \sin\theta_1\sin\theta_2
\ee
\eq{laplac} 
is the Laplacian for the transverse space. It is clear that the
solution of this is given by
\be
\label{soln}
f= A + \frac{B}{(r_1^2 + r_2^2)^3}
\ee
where $A$ and $B$  are constants.  

We now verify that the antisymmetric tensor equation of motion and
the Einstein equation also reduces to the Laplacian in transverse space
\eq{laplac}. The antisymmetric tensor equation is
\be
\partial_M(\sqrt{-g}e^{\Phi}H^{MNO}) = 2\kappa^2T_2\int
d^2\xi\epsilon^{i_1i_2}
\partial_{i_1}X^N\partial_{i_2}X^O\delta^{10}(x-X)
\ee
Substituting the ansatz in \eq{sugraein} the equation for the
anisymmetric tensor reduces to
\be
\partial_{r_1}(K\partial_{r_1} f) + \partial_{r_2} (K \partial_{r_2}
f) = -2T_2\kappa^2 K\frac{108}{(4\pi)^3}
\frac{\delta(r_1)}{r_1}\frac{\delta(r_2)}{r_2^5}
\ee
\eq{laplac} is identical to the above equation. Therefore the same
solution  \eq{soln} satisfies it.

The Einstein equation  is given in (3.15) of \cite{duff}.  
After some tedious but straight forward calculations
the components of Einstein equation along the
D1-brane reduce to 
\be
\frac{1}{2f^{7/4}} \partial_{r_1}(K\partial_{r_1}f) +
\frac{1}{2f^{7/4}}
\partial_{r_2} (K\partial_{r_2} f) = 
-T_2\kappa^2 K\frac{ 108 }{(4\pi)^3 f^{7/4}}
\frac{\delta(r_1)}{r_1}\frac{\delta(r_2)}{r_2^5}
\ee
This is the same as \eq{laplac}.
The remaining components of the Einstein equation reduce to identities
for the ansatz in \eq{sugraein}. The brane field equations are also
automatically satisfied.

To fix the constants $A$ and $B$ we go back to the string metric and
into the conventions of \cite{imsy}. $A=1$ as we require that at
infinity the metric reduce to that of the conifold. The constant $B$
by demanding that the net charge of the D1-branes is quantized. This
gives that 
\be
B= Ng_s\alpha^{\prime 3} \frac{864\pi^2}{16 +15\pi}
\ee
where $N$ is an integer.

\end{document}